\documentclass[prb,twocolumn,amssymb, nobibnotes, aps, superscriptaddress]{revtex4-2}
\usepackage{amsmath}
\usepackage{graphicx}
\usepackage{bm}
\usepackage{graphicx}
\usepackage{float}
\usepackage{subfigure}
\usepackage{makecell}
\usepackage{booktabs}
\usepackage{natbib}
\usepackage{color}
\usepackage[colorlinks=true, allcolors=blue]{hyperref}
\usepackage{wasysym}
\usepackage[utf8]{inputenc}

\usepackage{appendix}
\usepackage{changes}

\begin{document}
\title{Electronic properties and quantum transports in functionalized graphene Sierpinski carpet fractals}%
\author{Xiaotian Yang}
\affiliation{Key Laboratory of Artificial Micro- and Nano-structures of Ministry of Education and School of Physics and Technology, Wuhan University, Wuhan 430072, China}
\affiliation{Institute for Molecules and Materials, Radboud University, Heijendaalseweg 135, 6525 AJ Nijmegen, Netherlands}
\author{Weiqing Zhou}
\affiliation{Key Laboratory of Artificial Micro- and Nano-structures of Ministry of Education and School of Physics and Technology, Wuhan University, Wuhan 430072, China}
\affiliation{Institute for Molecules and Materials, Radboud University, Heijendaalseweg 135, 6525 AJ Nijmegen, Netherlands}
\author{Qi Yao}
\affiliation{Key Laboratory of Artificial Micro- and Nano-structures of Ministry of Education and School of Physics and Technology, Wuhan University, Wuhan 430072, China}
\affiliation{Institute for Molecules and Materials, Radboud University, Heijendaalseweg 135, 6525 AJ Nijmegen, Netherlands}
\author{Pengfei Lv}
\affiliation{Key Laboratory of Artificial Micro- and Nano-structures of Ministry of Education and School of Physics and Technology, Wuhan University, Wuhan 430072, China}
\author{Yunhua Wang}%
\email[Corresponding author: ]{wangyunhua@csrc.ac.cn}
\affiliation{Beijing Computational Science Research Center, Beijing, 100193, China}
\affiliation{Key Laboratory of Artificial Micro- and Nano-structures of Ministry of Education and School of Physics and Technology, Wuhan University, Wuhan 430072, China}
\affiliation{Institute for Molecules and Materials, Radboud University, Heijendaalseweg 135, 6525 AJ Nijmegen, Netherlands}
\author{Shengjun Yuan}%
\email[Corresponding author: ]{s.yuan@whu.edu.cn}
\affiliation{Key Laboratory of Artificial Micro- and Nano-structures of Ministry of Education and School of Physics and Technology, Wuhan University, Wuhan 430072, China}
\affiliation{Institute for Molecules and Materials, Radboud University, Heijendaalseweg 135, 6525 AJ Nijmegen, Netherlands}
\affiliation{Beijing Computational Science Research Center, Beijing, 100193, China}
\date{\today}
\begin{abstract}
Recent progress in controllable functionalization of graphene surfaces enables the experimental realization of complex functionalized graphene nanostructures, such as Sierpinski carpet (SC) fractals. Herein, we model the SC fractals formed by hydrogen and fluorine functionalized patterns on graphene surfaces, namely, H-SC and F-SC, respectively. We then reveal their electronic properties and quantum transport features. From calculated results of the total and local density of state, we find that states in H-SC and F-SC have two characteristics: (i) low-energy states inside about $|E/t|\leq 1$ (with $t$ as the near-neighbor hopping) are localized inside free graphene regions due to the insulating properties of functionalized graphene regions, and (ii) high-energy states in F-SC have two special energy ranges including $-2.3<E/t<-1.9$ with localized holes only inside free graphene areas and $3<E/t<3.7$ with localized electrons only inside fluorinated graphene areas. The two characteristics are further verified by the real-space distributions of normalized probability density. We analyze the fractal dimension of their quantum conductance spectra and find that conductance fluctuations in these structures follow the Hausdorff dimension. We calculate their optical conductivity and find that several additional conductivity peaks appear in high energy ranges due to the adsorbed H or F atoms.
\end{abstract}
\maketitle

\section{INTRODUCTION}

A fractal has a hierarchically self-similar block structure quantified by the non-integer Hausdorff dimension $d_{\rm H}$ \cite{geometryoffractal,gefen1984phase,pietronero2012fractals,feder2013fractals}. The unique self-similarity endows fractal nanostructures with a wealth of exotic and interesting physical features on electronic energy spectrum statistics \cite{iliasov2019power,iliasov2020linearized,kosior2017localization,hernando2015spectral}, quantum transport properties \cite{PhysRevB.94.161115,2016transport,opticalconductivity,han2019universal,bouzerar2020quantum,yang2020confined,iliasov2020hall,fremling2020existence}, 
plasmons \cite{westerhout2018plasmon}, flat bands \cite{nandy2015flat,nandy2015engineering,pal2018flat,nandy2021controlled} and topological phases \cite{pai2019topological,sarangi2021effect,fischer2021robustness,brzezinska2018topology,PhysRevResearch.2.023401}. Recently, nanoscale fractal structures, such as Sierpinski carpets (SC) and gaskets with atoms or molecules as building units, have been achieved by the bottom-up nanofabrication methods, including molecular self-assembly \cite{newkome2006nanoassembly,zhang2015controlling,jiang2017constructing,sun2015surface,nieckarz2016chiral,tait2015self,shang2015assembling}, chemical reactions \cite{zhang2016robust}, template packings \cite{li2017packing} and atomic manipulations in a scanning tunneling microscope \cite{kempkes2019design,PhysRevLett.126.176102,jiang2021direct}. In addition, SC nanostructures can also be created with arrays of waveguides \cite{xu2021shining}. Alternatively, top-down external field modulation is another feasible method for generating large-scale fractal structures. Especially, an external electric field is applied to two-dimensional materials so as to construct an electric-field-modulated SC fractal pattern \cite{yang2020confined}. However, the electrostatic fluctuation is induced by the gate method. Therefore, other external modulation manners generating robust fractal patterns and their transport properties need to be investigated.

In addition to its intrinsic Dirac physics, graphene is also an ideal engineering platform for searching new physical phenomena, because its 2D membrane surface is easy to be coupled with external electric and magnetic fields, mechanical tension and bending, atom vacancy and chemical functionalization. Especially, hydrogenated graphene \cite{elias2009control} and fluorinated graphene \cite{robinson2010properties,nair2010fluorographene} not only have high stability but also have considerable band gap with $3.5\sim5.7$ eV \cite{sofo2007graphane,PhysRevB.79.245117,cudazzo2010strong,pulci2010electronic} and $3.1\sim7.5$ eV \cite{zbovril2010graphene,karlicky2012band,bourlinos2012production,leenaerts2010first,samarakoon2011structural,csahin2011structures,yuan2015electronic}, respectively. Hydrogenated graphene has been produced by using atomic hydrogen beams \cite{guisinger2009exposure,sessi2009patterning,elias2009control} or exposure to hydrogen-based plasmas \cite{wojtaszek2011road,luo2009thickness}. Fluorinated graphene has been synthesized by direct gas-fluorination \cite{nair2010fluorographene,robinson2010properties}, plasma fluorination \cite{baraket2010functionalization,bon2009plasma}, hydrothermal fluorination \cite{gao2014effective,samanta2013highly} or photochemical/electrochemical methods \cite{lee2012selective}. Furthermore, functionalized regions in graphene membrane can be chemically controlled \cite{georgakilas2012functionalization,ferrari2015science,feng2016two,du2017broadband,ChengChang2021}. Therefore, it is highly possible to fabricate functionalized graphene SC fractal structures.

In this work, we model the SC fractals by using hydrogen and fluorine functionalization on the surface of graphene, namely, H-SC and F-SC, respectively. We investigate their electronic properties and quantum transports. Calculated results of the local and total density of states (LDOS and TDOS) indicate two remarkable characteristics: (i) at the low energy range about $|E/t|\leq 1$ ($t=2.8$ eV as the near-neighbor hopping), electrons and holes are confined inside free graphene regions because of the insulating properties of functionalized graphene parts; and (ii) at higher energy, in F-SC there exist two energy ranges, including $-2.3<E/t<-1.9$ where holes are localized only inside free graphene areas and $3<E/t<3.7$ where electrons are localized only inside fluorinated graphene areas. Analyses on real-space distributions of normalized probability density further verify the two characteristics. We discuss the fractal dimension of their conductance spectra and find that conductance fluctuations follow the Hausdorff fractal dimension. In these functionalized graphene SC fractals, additional multiple optical conductivity peaks appear in high energy ranges due to the adsorbed H or F atoms.

%%%%%%%%%%%%%%%%%%%%%%%%%%%%%%%%% FIG 1 %%%%%%%%%%%%%%%%%%%%%%%%
\begin{figure}[H]
\centering
\includegraphics[width=10.8cm]{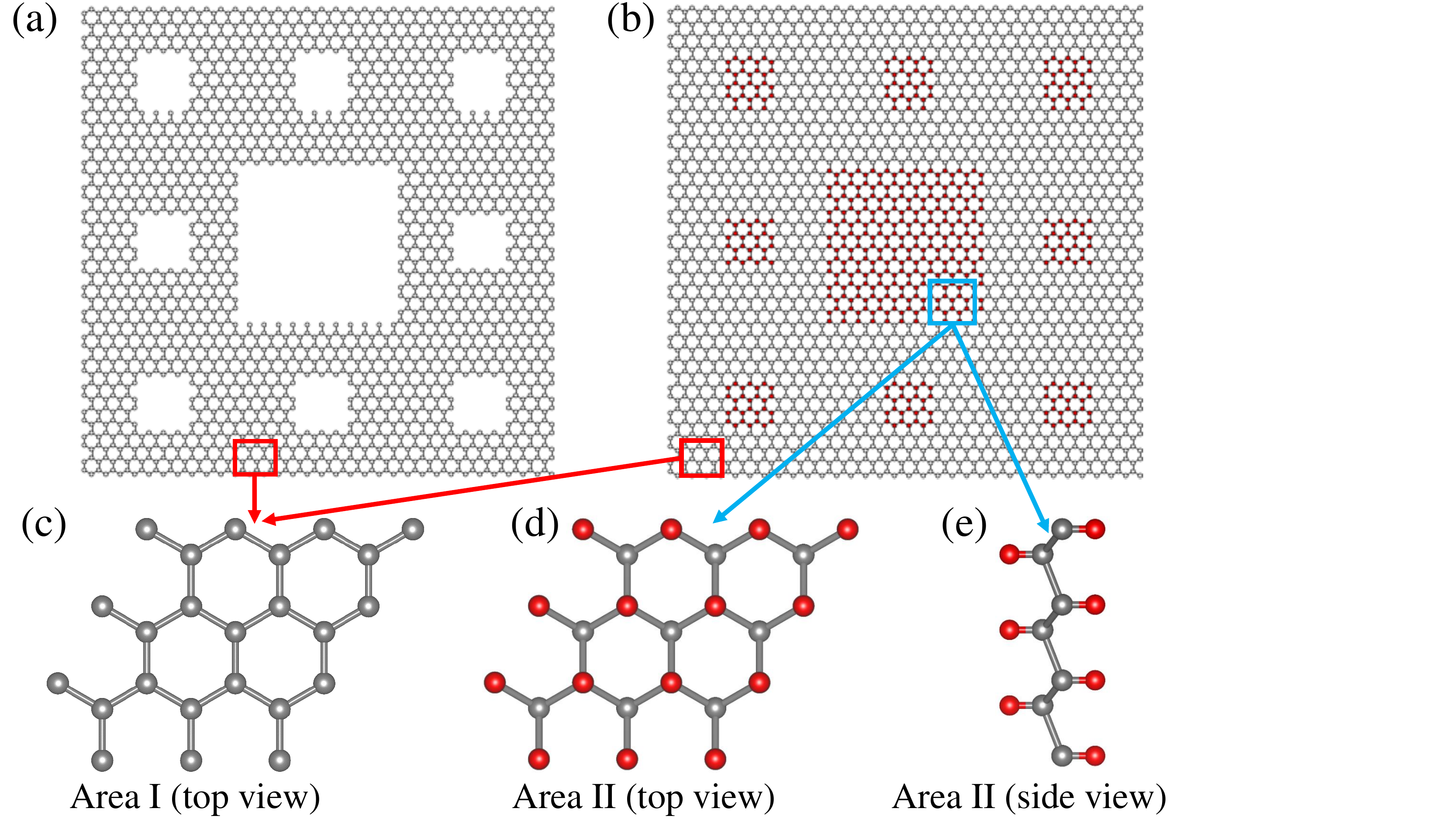}
    \caption{An illustration of graphene-based SCs. (a) A SC sample generated by atom vacancies with the iteration number $I =2$ and the square width $W=32.5a$ with $a$ as the lattice constant of graphene. (b) A schematic of the functionalized SC by the surface adatoms in corresponding zones, i.e., Area II with hydrogen or fluorine denoted by red color balls. The zoom-in figures show spatial lattice structures inside Area I in (c) and inside Area II from top and side views in (d) and (e), respectively.}\label{sample}
\end{figure}
%%%%%%%%%%%%%%%%%%%%%%%%%%%%%%%%% FIG 1 %%%%%%%%%%%%%%%%%%%%%%%%

\section{MODEL AND METHODS}\label{sec:model}

We investigate two kinds of graphene-based SC structures, including the SC generated by atom vacancies in Fig.~\ref{sample}(a) and the SC functionalized by hydrogen or fluorine in Fig.~\ref{sample}(b). For conveniences in later discussions, we denote the SC formed by atom vacancies as A-SC, and we name the SC formed by hydrogenated- and fluorinated-graphene as H-SC and F-SC, respectively. The functionalized SC structures generally have partly out-of-plane distortions induced by adsorbed atoms \cite{yuan2015electronic}, for instance, as presented in Figs.~\ref{sample}(d) and~\ref{sample}(e), where $I =2$ is the iteration number and $W=32.5a$ is the square width with $a$ as the lattice constant of graphene. Electrons in these graphene-based SCs are described by the following tight-binding Hamiltonian:
\begin{equation}
\begin{aligned}
H = \sum_{<ij> \alpha \beta}t^{ij}_{\alpha\beta}c^{\dag}_{i,\alpha}c_{j,\beta}+ \sum_{i \alpha}\varepsilon^{i}_{\alpha}n_{i,\alpha},
\end{aligned}\label{equation1}
\end{equation}
where $i$, $j$ are site indices, $\langle ij \rangle$ represents nearest-neighbor sites, atomic orbitals are marked with notation $\alpha$ and $\beta$, $c^{\dag}_{i,\alpha}$ ($c_{i,\alpha}$) creates (annihilates) fermions with orbital $\alpha$ at site $i$. Electron hopping between various sites is described in the first term, and the second term gives the on-site potential. When SC changes from the $I$th iteration to the $(I+1)$th iteration, the unit is replicated with ${\cal N}=8$ times larger in area and ${\cal L}=3$ times larger in width. The Hausdorff dimension is defined by $d_{\rm H}  \equiv \log_{{\cal L}}{\cal N}$ $\simeq 1.89$.

The distinctions between H-SC and F-SC mainly lie on the spatially structural distortion and the types of orbitals. Considering that the hybridization between hydrogen and carbon atoms involves only $1s$ orbital of hydrogen and $2p_{z}$ orbital of carbon, we adopt the $\pi$-band TB model for H-SC. The nearest-neighbor hopping between carbon atoms is labeled by $t^{ij}_{p_{z}p_{z}}=t$, the nearest-neighbor hopping between carbon and hydrogen is $t^{ij}_{p_{z}s}=2t$, and the on-site potentials are $\varepsilon^{i}_{p_{z}}=0$ and $\varepsilon^{i}_{s}=-t/16$ \cite{PhysRevLett.105.056802,TBPM}. However, the hybridization between fluorine and carbon atoms involves multiple orbitals, including $2s$, $2p_{x}$, $2p_{y}$ and $2p_{z}$ orbitals from carbon and $2p_{x}$, $2p_{y}$ and $2p_{z}$ orbitals from fluorine. We use the multiple-orbital TB model for F-SC and adopt the parameters obtained from the \textit{ab initio} density-functional theory calculations (see the details in Appendix \ref{sec:append_A}) \cite{yuan2015electronic}.

We use the tight-binding propagation method (TBPM) to calculate the density of states (DOS). We start the evolution of a quantum system with a random initial state $|\varphi(0)\rangle$, which is normalized superposition of all basis states $\sum_{n}A_{n}|n\rangle$. The DOS is calculated via Fourier transform of the correlation function \cite{hams2000fast,TBPM}:
\begin{equation}
\begin{aligned}
D(E)=\frac{1}{2\pi}\int_{-\infty}^{\infty}e^{iEt}\langle\varphi(0)|e^{-iHt}|\varphi(0)\rangle dt.
\end{aligned}
\label{equation2}
\end{equation}
After the Fourier transform of states at different time during the evolution $|\varphi(t)\rangle = e^{-iHt}|\varphi(0)\rangle$, we obtain the quasi-eigenstates $|\psi(E)\rangle$ by \cite{kosloff1983fourier,TBPM}
\begin{equation}
\begin{aligned}
|\psi(E)\rangle&=\frac{1}{2\pi}\int_{-\infty}^{\infty}dte^{iEt}|\varphi(t)\rangle\\&=\frac{1}{2\pi}\sum_{n}A_{n}\int_{-\infty}^{\infty}dte^{i(E-E_{n})t}|n\rangle\\&=\sum_{n}A_{n}\delta(E-E_{n})|n\rangle,
\end{aligned}
\label{equation3}
\end{equation}
which can be further normalized as
\begin{equation}
\begin{aligned}
|\psi(E)\rangle=\frac{1}{\sqrt{\sum_{n}|A_{n}|^{2}\delta(E-E_{n})}}\sum_{n}A_{n}\delta(E-E_{n})|n\rangle.
\end{aligned}
\label{equation4}
\end{equation}
For the finite fractal structure, one can make an average by different realizations of random coefficients $A_{n}$ to obtain more accurate results of $D(E)$ and $|\psi(E)\rangle$.

%%%%%%%%%%%%%%%%%%%%%%%%%%%%%%%%% FIG 2 %%%%%%%%%%%%%%%%%%%%%%%%
\begin{figure*}[tbp]
\centering
\includegraphics[width=1.0\textwidth]{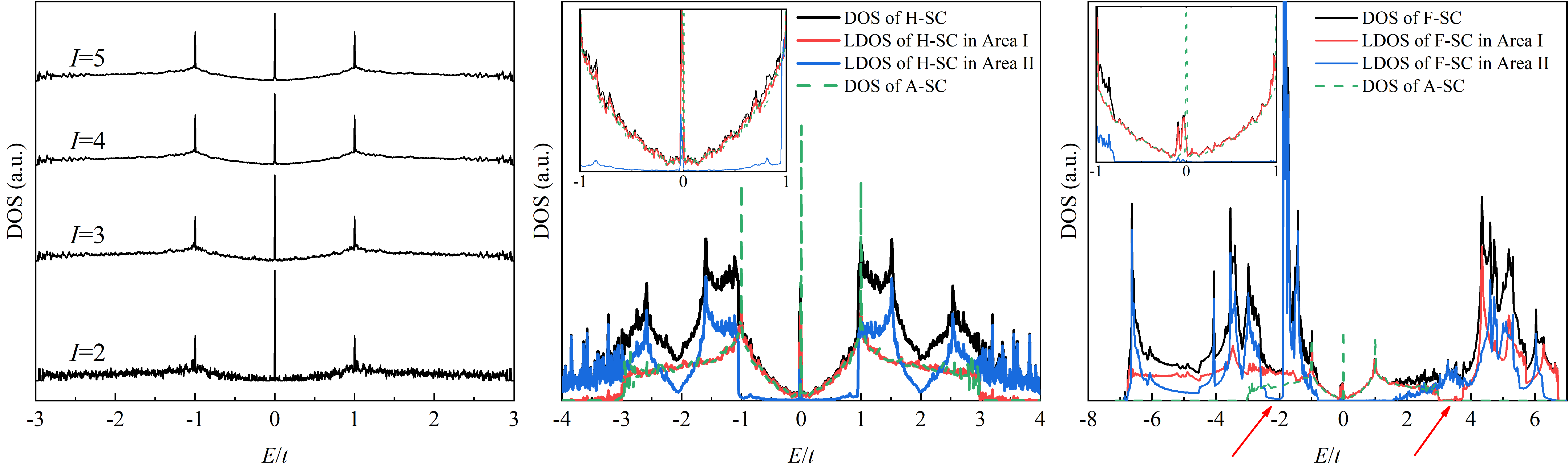}
\caption{(a) DOS of A-SC for various iterations $I$ ranging from two to five. (b) DOS and LDOS of H-SC with $I=4$ and $W=297.5 a$. (c) DOS of F-SC with $I=4$ and $W=297.5 a$. In (b) and (c) the DOS of A-SC is added for comparisons.} 
\label{DOS_converged}
\end{figure*}
%%%%%%%%%%%%%%%%%%%%%%%%%%%%%%%%% FIG 2 %%%%%%%%%%%%%%%%%%%%%%%%

We use the Kubo formula to obtain the real part ${\rm Re} [\sigma_{\alpha\beta}(\omega)]$ of the optical conductivity \cite{TBPM,kubo1957statistical,statisticalphysics}, as follows: 
\begin{equation}
\begin{split}
{\rm Re} [\sigma_{\alpha\beta}(\omega)]=&\lim_{\epsilon \to 0^{+}}\frac{e^{-\hbar\omega/k_{B}T}-1}{\hbar\omega\Omega}\int_{0}^{\infty}e^{-\varepsilon t}sin\omega t \\ &\times [2{\rm Im}\langle\varphi_{2}(t)|J_{\alpha}|\varphi_{1}(t)\rangle_{\beta}] dt,
\end{split}\label{equation5}
\end{equation}
where $\alpha, \beta = x, y$, $\Omega$ is the area of SC sample, and the current operator is written as $J_{\alpha}=-\frac{ie}{\hbar}\sum_{i,j}t_{ij}(\mathbf{r}_{j}-\mathbf{r}_{i})_{\alpha}c^{\dag}_{i}c_{j}$. Here, $\varphi_{1}(t)$ and $\varphi_{2}(t)$ have following definitions~\cite{PhysRevE.56.1222,TBPM}:
\begin{gather}
|\varphi_{1}(t)\rangle_{x}=e^{-iHt/\hbar}[1-f(H)]J_{x}|\varphi\rangle,\\
|\varphi_{1}(t)\rangle_{y}=e^{-iHt/\hbar}[1-f(H)]J_{y}|\varphi\rangle,\\
|\varphi_{2}(t)\rangle=e^{-iHt/\hbar}f(H)|\varphi\rangle,
\label{equation8}
\end{gather}
where $f(H)=\frac{1}{e^{\beta(H-\mu)}+1}$ is the Fermi-Dirac distribution operator.

The quantum conductance as an auxiliary quantity with the help of box-counting analysis \cite{guarneri2001fractal} is calculated by the Landauer formula \cite{groth2014kwant},
\begin{equation}
\begin{split}
G_{ab}=\frac{e^2}{h}\sum_{i\in{a},j\in{b}}|S_{ij}|^2,
\end{split}
\label{equation9}
\end{equation}
where $S_{ij}$ is the scattering matrix with $a$ and $b$ denoting two electrodes.

\section{RESULTS AND DISCUSSION}\label{sec:result}

In this section, we discuss DOS, quantum conductance and optical conductivity of A-SC, H-SC and F-SC structures and make comparisons with each other. Firstly, we analyze the TDOS and LDOS in the three structures. Then, we calculate quasi-eigenstates and discuss their real-space  probability density distribution. Next, we calculate the quantum conductance of A-SC and H-SC structures without considering the lattice vibration and compare their results with that of the SC formed by a local electric field \cite{yang2020confined}. We further estimate the fractal dimension of their conductance spectrum by the box-counting algorithm \cite{guarneri2001fractal}. At last, we discuss the optical absorption properties of these considered structures and compare them with each other.

%%%%%%%%%%%%%%%%%%%%%%%%%%%%%%%%% FIG 3 %%%%%%%%%%%%%%%%%%%%%%%%
\begin{figure*}[tbp]
\centering
\includegraphics[width=0.93\textwidth]{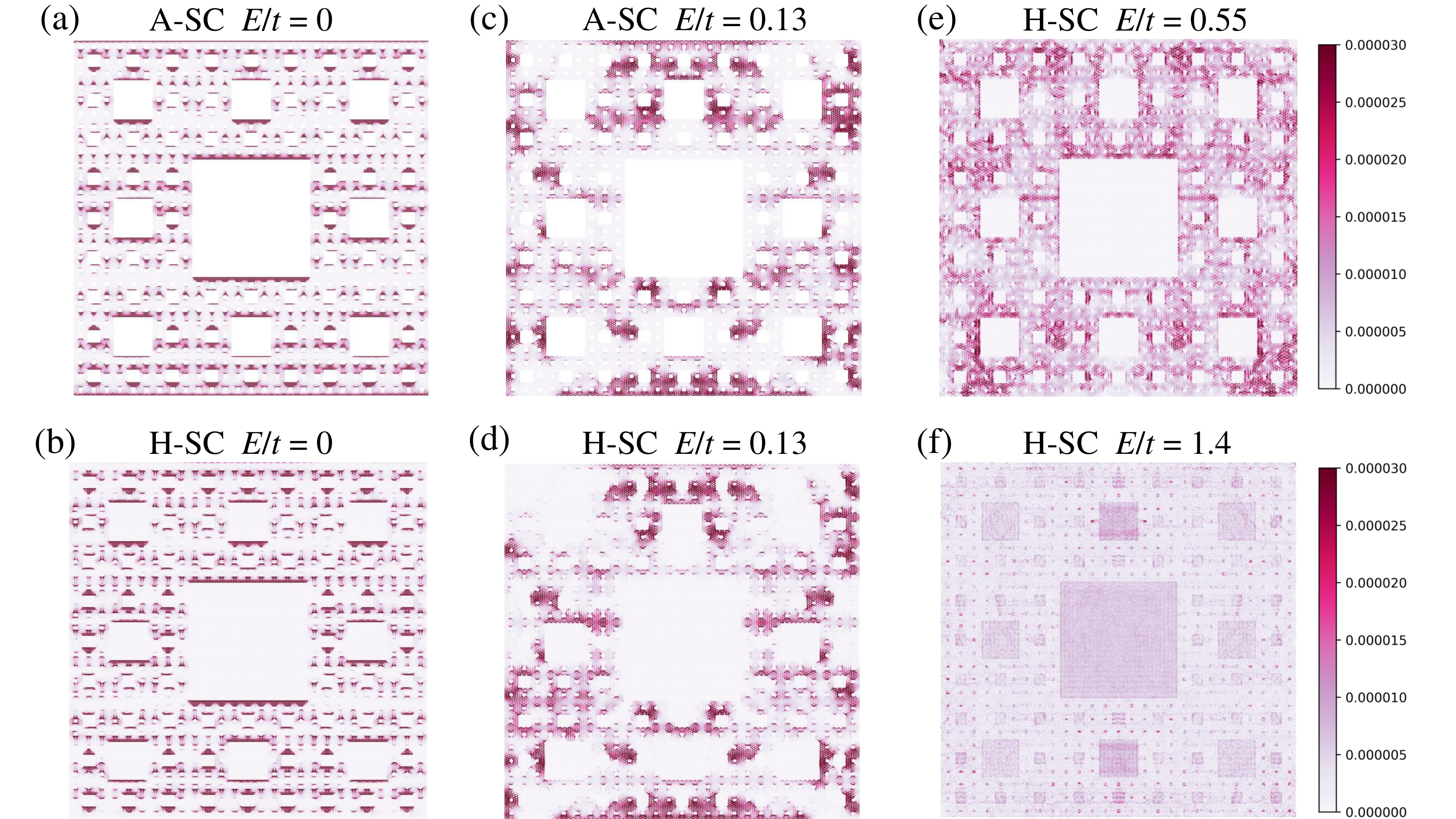}
\caption{The real-space distributions of normalized probability density of states at $E/t=0$ for A-SC in (a) and H-SC in (b), at $E/t=0.13$ for A-SC in (c) and H-SC in (d), and at $E/t=0.55$ in (e) for H-SC and $E/t=1.4$ in (f) for H-SC. Here $I = 4$ and $W = 297.5 a$ for both A-SC and H-SC.}
\label{quasi_H-SC}
\end{figure*}
%%%%%%%%%%%%%%%%%%%%%%%%%%%%%%%%% FIG 3 %%%%%%%%%%%%%%%%%%%%%%%%

\subsection{DOS}

Figure~\ref{DOS_converged}(a) shows the DOS of A-SC as a function of energy under various iteration numbers $I$ ranging from $2$ to $5$. As we can see, the DOS relatively changes little after $I\geq4$. Hence, we mainly choose $I=4$ to produce these SC structures so as to later explore their electronic and transport features in these exotic fractal systems. For the DOS of A-SC in Fig.~\ref{DOS_converged}(a), we can observe two characteristics: (i) two sharp peaks appear at $E/t = \pm1$, which are the Van Hove singularities similar to those of pristine graphene, and (ii) an additional central peak exists at $E/t = 0$, which is absent in pristine graphene and reflects the edge states induced by the zigzag terminations in the A-SC structure.

Using Eq.~\eqref{equation2}, we further calculate the DOS of H-SC and F-SC fractal structures with $I=4$ and plot the results in Figs.~\ref{DOS_converged}(b) and~\ref{DOS_converged}(c), respectively. Inside the low energy range about $-1\leq E/t\leq 1$, we can clearly see the DOS is mainly contributed by LDOS of Area I in both H-SC and F-SC structures. In addition, the DOS of H-SC and F-SC are almost the same as that of A-SC except a little deviation of the DOS of F-SC near $E/t \sim -1$, because some states appear into fluorinated Area II, as shown in the inset in Fig.~\ref{DOS_converged}(c). These compared results indicate that even though various orbitals of C and F atoms are taken into account, the states inside low energy regions about $-1\leq E/t\leq 1$ are contributed by the $p_z$ orbital of C atoms in these fractal structures. In addition, these compared results reflect the fact that, inside low energy range these states in both H-SC and F-SC fractal structures are confined inside Area I. In other words, hydrogenated or fluorinated Area II in Figs.~\ref{DOS_converged}(b) and~\ref{DOS_converged}(c) plays the same role as atom vacancies in Fig.~\ref{DOS_converged}(a) for electrons inside low energy range. These results can also be explained by the insulating features of fully hydrogenated and fluorinated graphene \cite{TBPM, yuan2015electronic}, which forbid the penetration of low energy states into Area II. In this respect, we say that H-SC and F-SC are effective analogues of A-SC. For less perfect adatom positions with weak defect densities $\rho$ less than $3\%$, the confinement of low-energy states inside Area I can be remained (see Appendix \ref{sec:append_B}). We figure out that these dangling $s$, $p_x$ and $p_y$ orbitals of C atoms at edges must be saturated if a multiple-orbital TB model is adopted for the A-SC structure, otherwise, unreasonable DOS results will be given inside low energy regions. In F-SC, these inner orbitals are naturally saturated by $p$ orbitals of F atoms. Consequently, the $p_z$ orbital TB model effectively captures the low energy physics in these fractal structures.

At higher energy ranges in Figs.~\ref{DOS_converged}(b) and~\ref{DOS_converged}(c), from the LDOS in Area II we can see that states apparently exist in Area II for both H-SC and F-SC structures. These compared results between the DOS of functionalized SC structures and the DOS of the A-SC structure indicate that these states are also contributed by the other orbitals from hydrogen or fluorine inside higher energy regions at $|E/t|>1$. For H-SC in Fig.~\ref{DOS_converged}(b), we can see that additional Van Hove singularities appear at about $|E/t| = 1.5$ and $2.5$, as well as other higher energy positions. For F-SC in Fig.~\ref{DOS_converged}(c), various additional Van Hove singularities emerge at several higher energy positions. The electron-hole symmetry is obviously broken by $p$ orbitals of fluorine and $p_{x}$ and $p_{y}$ orbitals of carbon. This is different from that in H-SC, where the spherical $s$ orbital of hydrogen remains the electron-hole symmetry. As a result, inside higher energy regions, electronic states will exhibit different behaviors in functionalized SC structures compared with that in A-SC structure, such as state distributions in real space and optical absorptions, which will be discussed later. In addition, we can see two special high energy ranges denoted by two arrows in F-SC: (i) at about $-2.3<E/t<-1.9$, where holes are mainly located inside Area I, and (ii) at about $3<E/t<3.7$, where electrons are mainly located inside Area II. This means that we can choose energy windows to localize holes and electrons in corresponding Area I and Area II in F-SC fractal structure.

\subsection{Quasi-eigenstates}

Figure~\ref{quasi_H-SC} presents the real-space distribution of normalized probability density of states in A-SC and H-SC fractal structures. We first discuss the probability density distributions of zero-energy states in Fig.~\ref{quasi_H-SC}(a) for A-SC and in Fig.~\ref{quasi_H-SC}(b) for H-SC. Obviously, these nonzero probability densities are mainly located at the zigzag terminations. Therefore, the central peaks at zero energy in Figs.~\ref{DOS_converged}(a) and ~\ref{DOS_converged}(b) correspond to the edge states \cite{yang2020confined}. For the low energy states in Figs.~\ref{quasi_H-SC}(c)-\ref{quasi_H-SC}(e), these nonzero probability densities are localized inside Area I. Some states in H-SC and A-SC can possibly exhibit very similar probability density distributions, for instance, the states at $E/t=0.13$ in Figs.~\ref{quasi_H-SC}(c) and~\ref{quasi_H-SC}(d). It is a fact that an arbitrary state in A-SC structure can impossibly appear inside these regions of atom vacancies. However, these states with higher energy can exist in functionalized SC structures, as presented in Figs.~\ref{DOS_converged}(b) and~\ref{DOS_converged}(c). We also calculate the probability density distribution of high-energy states in H-SC, for instance, the state at $E/t=1.4$, as shown in Fig.~\ref{quasi_H-SC}(f). There are apparent nonzero probability densities inside Area II, and hence these high-energy extended states in H-SC are remarkably different from those in A-SC. In this situation, H-SC can not be viewed as an effective fractal of A-SC.

%%%%%%%%%%%%%%%%%%%%%%%%%%%%%%%%% FIG 4 %%%%%%%%%%%%%%%%%%%%%%%%
\begin{figure*}[tbp]
\centering
\includegraphics[width=0.93\textwidth]{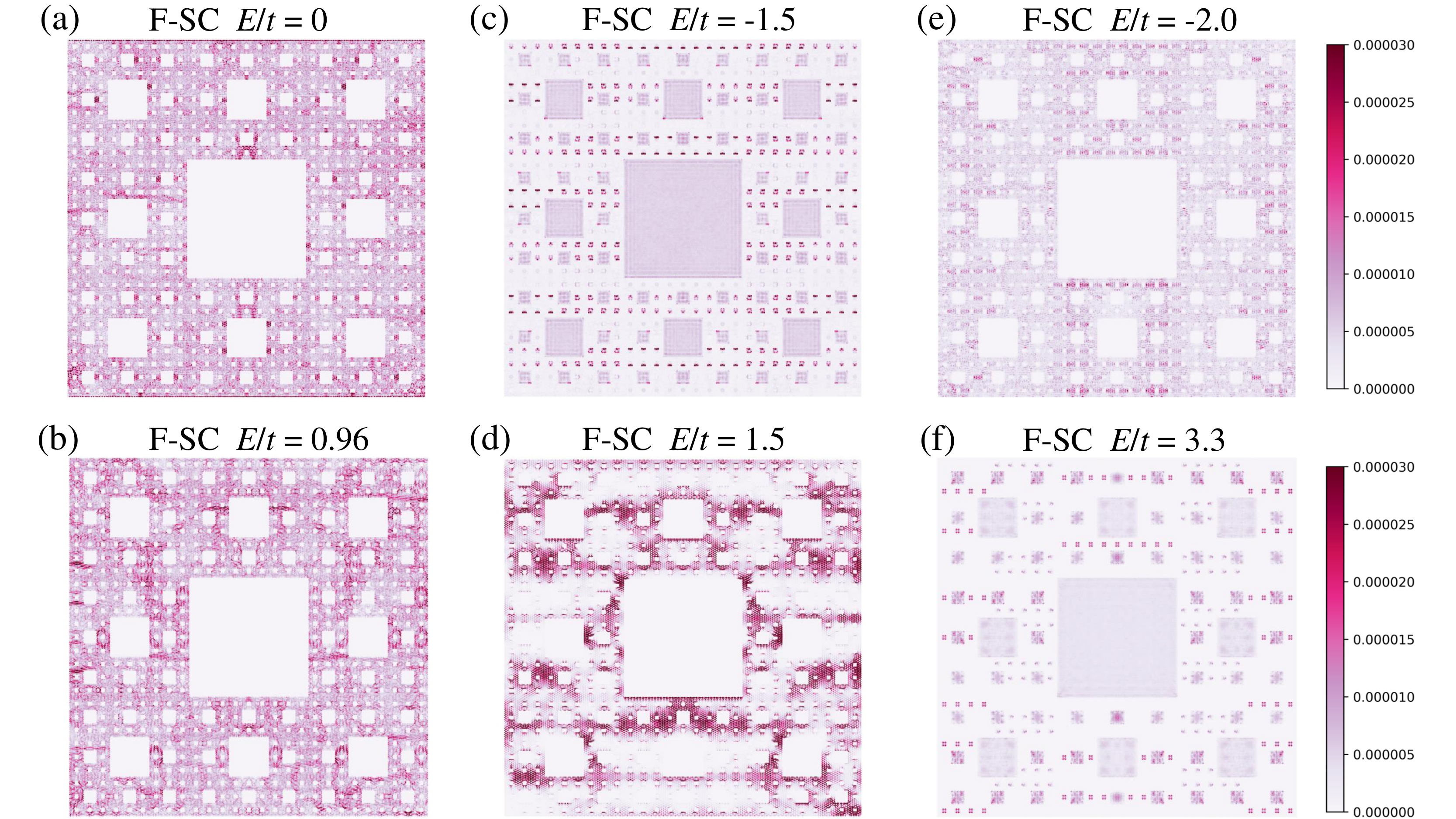}
\caption{The real-space distributions of normalized probability density of states for H-SC at $E/t=0$ in (a), $E/t=0.96$ in (b), $E/t=-1.5$ in (c), $E/t=1.5$ in (d), $E/t=-2$ in (e), and $E/t=3.3$ in (f). Here $I = 4$ and $W = 297.5 a$ for F-SC.}\label{quasi_F-SC}
\end{figure*}
%%%%%%%%%%%%%%%%%%%%%%%%%%%%%%%%% FIG 4 %%%%%%%%%%%%%%%%%%%%%%%%

For the F-SC fractal structure, the real-space distributions of normalized probability density of states at various energy are shown in Fig.~\ref{quasi_F-SC}. Because the zero-energy state in F-SC has a broadening compared with that of A-SC [see the inset of Fig.~\ref{DOS_converged}(c)], the nonzero probability densities of the zero-energy state appear not only at structural edges but also inside Area I, as shown in Fig.~\ref{quasi_F-SC}(a). For low energy regions, even at $E/t=0.96$ in Fig.~\ref{quasi_F-SC}(b), electrons are still localized inside Area I, due to the wider band gap of fluorinated graphene \cite{yuan2015electronic}. To present the electron-hole asymmetry in F-SC, as an example, we calculate the probability densities of the states at $E/t=-1.5$ and $E/t=1.5$ and plot the results in Fig.~\ref{quasi_F-SC}(c) and \ref{quasi_F-SC}(d), respectively. We can see that the states at $E/t=-1.5$ are located inside fluorinated Area II, but the states at $E/t=1.5$ are still mainly inside Area I. At last, we check the states inside the two energy windows in which hole and electron
are respectively localized as shown in previous Fig.~\ref{DOS_converged}(c). We plot the real-space distributions of the normalized probability density for states at $E/t=-2.0$ and $E/t=3.3$ in Fig.~\ref{quasi_F-SC}(e) and \ref{quasi_F-SC}(f), respectively. Indeed, inside the two energy windows, the hole state is localized well inside Area I and the electron state is localized inside Area II.

\subsection{Conductance fluctuations}

Generally, the fractal dimension is embodied in quantum transports. Consequently, there is a close correlation between conductance fluctuations and the geometrical dimension of fractals \cite{2016transport}. In addition, conductance fluctuations are also observed in generic chaotic cavities \cite{ketzmerick1996fractal}, quantum billiards \cite{crook2003imaging}, quasi-ballistic gold nanowires \cite{hegger1996fractal}, and diffusive/ballistic semiconductor devices \cite{marlow2006unified}. For a fractal such as SC with an infinite ramification number \cite{gefen1984phase}, the box-counting dimension of the conductance fluctuations is generally equal to the Hausdorff dimension, which has been checked theoretically \cite {2016transport}. Besides, this correlation is still approximately valid for a fractal with a finite but relatively large ramification number, namely, $I\geq 4$ \cite{2016transport,yang2020confined}.

%%%%%%%%%%%%%%%%%%%%%%%%%%%%%%%%% FIG 5 %%%%%%%%%%%%%%%%%%%%%%%%
\begin{figure*}[tbp]
\centering
\includegraphics[width=0.93\textwidth]{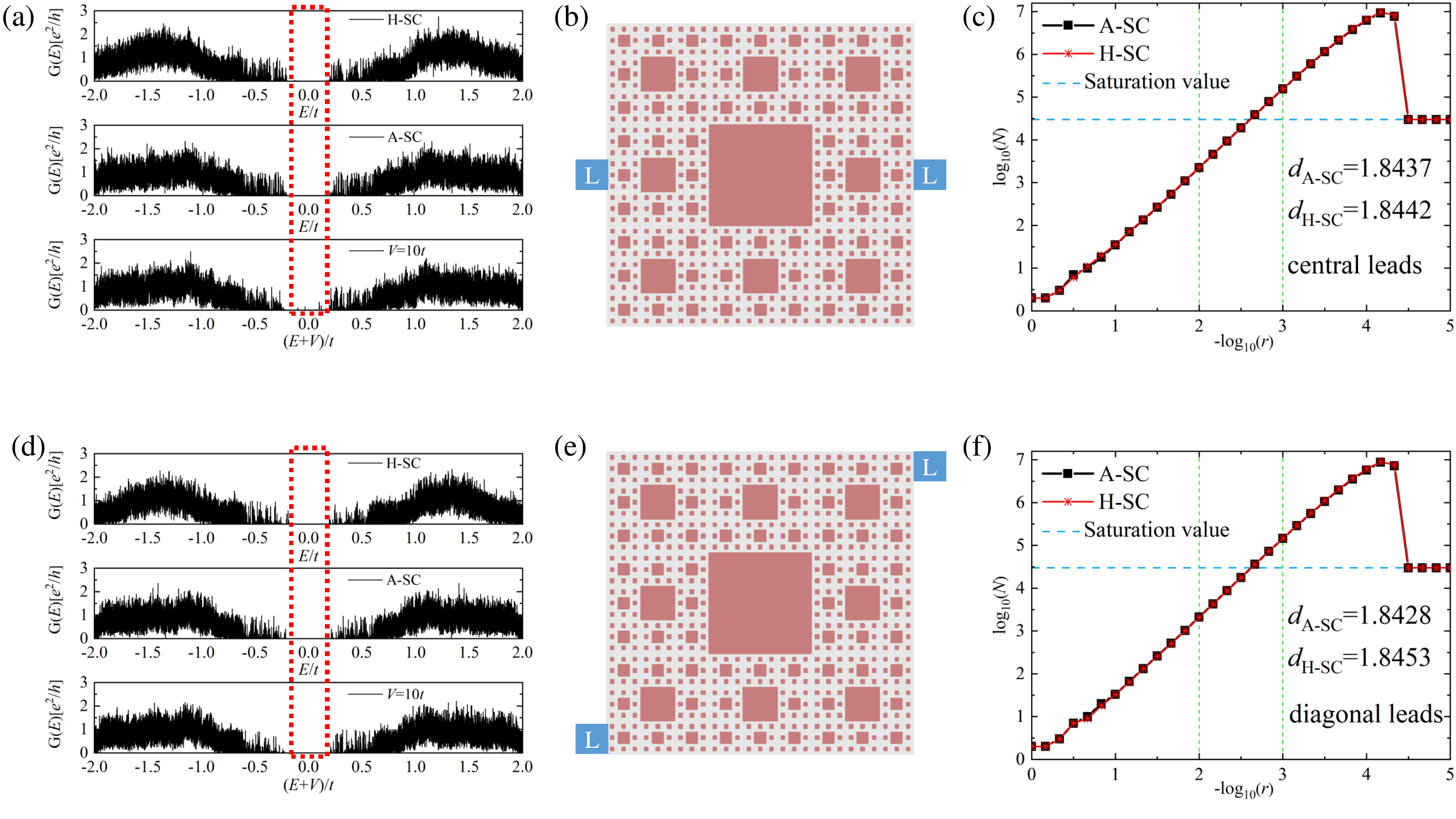}
\caption{(a, d) Conductance $G(E)$ in units of $e^{2}/h$ as a function of energy $E$ in H-SC, A-SC, and the SC formed by applying an external electric field with $V=10t$, where $I=4$ and $W=297.5 a$. The lead (L) configurations are correspondingly at (middle/middle, bottom/top) positions as sketched in (b, e), where lead width is $(3+\frac{2\sqrt{3}}{3})a$. (c, f) Box-counting algorithm analysis of the conductance fluctuations for (a, d).}\label{conduc_BC}
\end{figure*}
%%%%%%%%%%%%%%%%%%%%%%%%%%%%%%%%% FIG 5 %%%%%%%%%%%%%%%%%%%%%%%%

Detailed calculations on the quantum conductance are implemented in KWANT \cite{groth2014kwant} by Eq.~\eqref{equation9}. Here we discuss conductance fluctuations in three fractal structures, including H-SC, A-SC and the SC formed by local electric field modulation with $V=10t$ \cite{yang2020confined}. The conductance calculation of F-SC is not performed since the matrix dimension in F-SC for $I =4$ is beyond the calculation limit in KWANT.

Figure~\ref{conduc_BC}(a) shows the calculated conductance $G(E)$ as a function of energy $E$ for the considered three structures with $I =4$ and $W=297.5a$, where the middle lead configurations are shown in Fig.~\ref{conduc_BC}(b). We can see a remarkable conductance gap of $G(E)$ where the conductance vanishes in Fig.~\ref{conduc_BC}(a). The conductance gap is a hallmark of electronic transports in graphene-based fractal structures \cite{2016transport, yang2020confined}. There are several minor conductance peaks inside the conductance gap in the SC formed by a local electric field even with large potential magnitude of $V=10t$. Therefore, we should figure out that the SC generated by a local electric field ($V=10t$) is not a perfect equivalent fractal of A-SC even inside low energy regions. However, both H-SC and A-SC exhibit a perfect conductance gap in low energy regions.

When energy $E$ stays away from the central area, the conductance spectrum contains many fluctuations, making the conductance curve quite noisy. The fluctuations can be characterized by the dimension of the whole conductance spectrum. The dimension of the conductance spectrum is obtained by the box-counting algorithm \cite{guarneri2001fractal}. We count the number $N$ of the box of different size $r$ such that the box continuously and completely covers the graph of conductance for H-SC and A-SC in Fig.~\ref{conduc_BC}(a). The box-counting results of the log value of counting number $N$ as a function of the minus log value of box size $r$ are plotted in Fig.~\ref{conduc_BC}(c). For large values of $r$, where $-\log_{10}(r)$ is around 0, the box is too large to grasp the features of the conductance fluctuations. For very small $r$ below $\sim 10^{-4}$, i.e., $-\log_{10}(r) \geq 4$, each box covers only one data point due to very small size so that $N$ is not increased anymore but turns to be invariant, namely, saturated. There is an intermediate $r$ called as the "scaling region", where the scaling between $\log_{10}N(-\log_{10}(r))$ and $-\log_{10}(r)$ is linear in the plot. The slope of $\log_{10}N(-\log_{10}(r))$ as a function of $-\log_{10}(r)$ inside the "scaling region" from $2$ to $3$, i.e., the box-counting dimension, is extracted, as shown in Fig.~\ref{conduc_BC}(c). The extracted values of box-counting dimension are $d_{\rm H-SC}=1.8453$ and $d_{\rm A-SC}=1.8428$ for H-SC and A-SC, respectively. The Hausdorff dimension is given by $d_{\rm H} = 1.89$ for our considered SC fractals. Our results in A-SC and H-SC with $I=4$ are close to the Hausdorff dimension $d_{\rm H}$. We might possibly further infer that the slight difference between the box-counting dimension and Hausdorff dimension will vanish if the ramification number is infinite.

We also consider the position effects of leads on the conductance spectrum and box-counting dimension in Figs.~\ref{conduc_BC}(d)-\ref{conduc_BC}(e). We can see similar conductance gaps and fluctuations. The box-counting dimension results change little. Consequently, we think that the lead position has a weak influence on transport features of A-SC and H-SC with a relative large ramification number.

\subsection{Optical conductivity}
\begin{figure}[H]
\centering
\includegraphics[width=8.5cm]{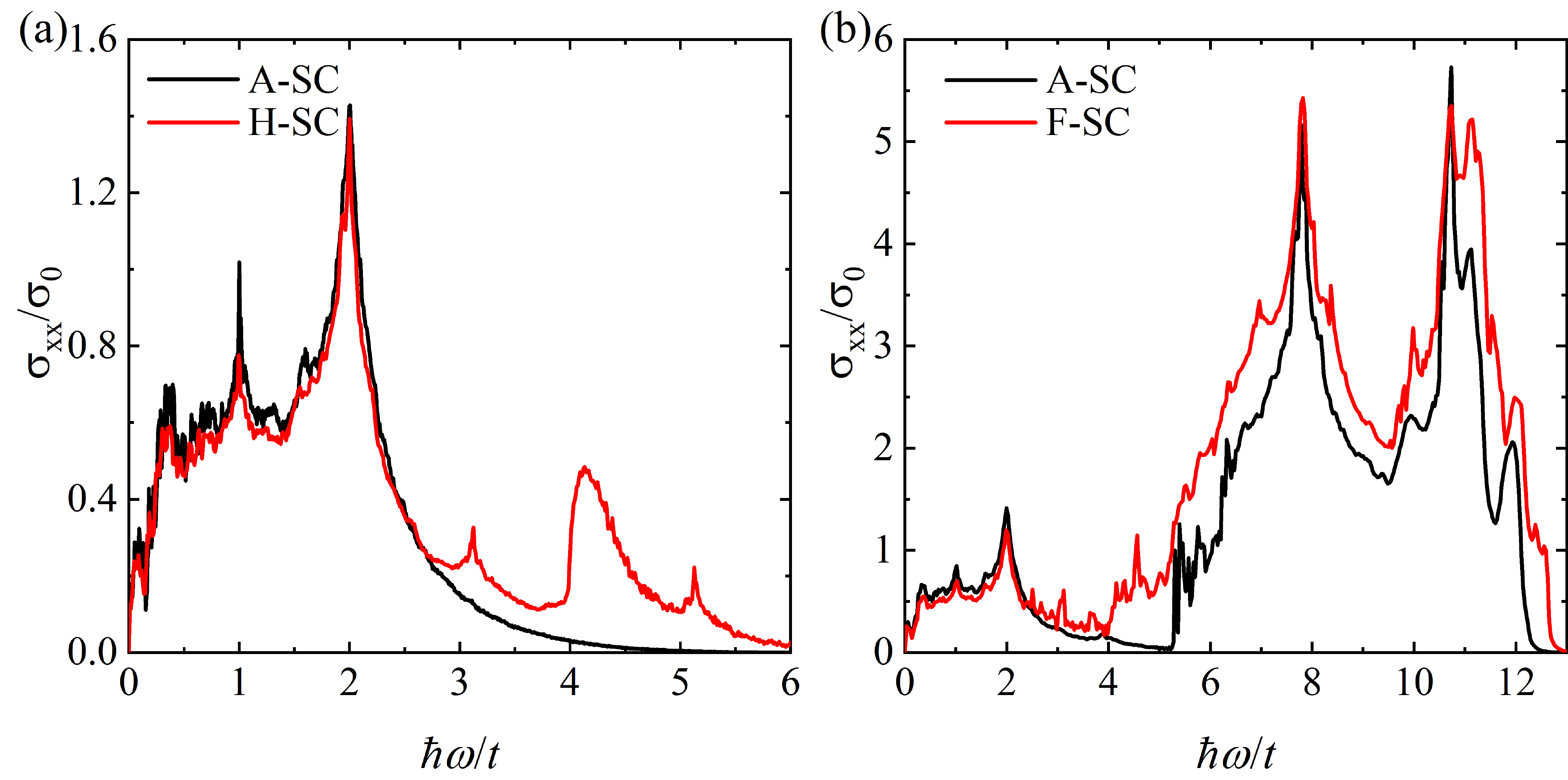}
\caption{Real part of optical conductivity $\sigma_{xx}$ of H-SC in (a) and F-SC in (b) as a function of photon energy $\hbar\omega/t$ at $\mu=0$ and $T=300$ K. The $\sigma_{xx}$ of A-SC is also added for comparisons. Here $I=4$ and $W=297.5a$ for all SC structures.}
\label{AC_conductivity}
\end{figure}

We further investigate the optical conductivity of the A-SC, H-SC and F-SC fractal structures. Using Eqs.~\eqref{equation5}-\eqref{equation8}, we calculate the optical conductivity $\sigma_{xx}$ as a function of $\omega/t$ and plot the real part of optical conductivity in Fig.~\ref{AC_conductivity}, which corresponds to the optical absorption. When $\omega/t$ is about below $2.1$, A-SC, H-SC and F-SC exhibit very close optical conductivity behaviors. Except for additional few peaks around $\omega/t\sim 1$ the almost constant value of the optical conductivity inside low frequency is similar to the universal value of pristine graphene \cite{TBPM}. The LDOS of Area II in H-SC and F-SC has a band gap of about $2t$ [see Figs.~\ref{DOS_converged}(b) and (c)], which leads to zero optical conductivity when the $\omega$ is below 2t. Thus, the optical properties of H-SC and F-SC at low frequency range are actually dominated by the clean graphene in Area I.

However, various absorption peaks in H-SC and F-SC structures appear inside a higher frequency range. For instance, the additional three peaks appear around $\omega/t=3.1$, $4.1$ and $5.1$ in H-SC due to the adsorbed H atoms. For F-SC, the optical conductivity at higher frequency range exhibit several strong absorption peaks, which are also predicted in fully fluorinated graphene \cite{yuan2015electronic}. In other words, functionalized F-SC fractal structures interestingly inherit the low frequency optical properties of pristine graphene and the higher frequency optical properties of fully fluorinated graphene.

%%%%%%%%%%%%%%%%%%%%%%%%%%%%%%%%% FIG 6 %%%%%%%%%%%%%%%%%%%%%%%%
%%%%%%%%%%%%%%%%%%%%%%%%%%%%%%%%% FIG 6 %%%%%%%%%%%%%%%%%%%%%%%%

\section{SUMMARY}\label{sec:summary}

In summary, we investigated the DOS, real-space distributions of normalized probability density, conductance fluctuations and optical conductivity of H-SC and F-SC fractal structures. Our DOS results show that low-energy states with $|E/t|\leq 1$ in H-SC and F-SC are mainly located inside free graphene regions because of the insulating properties of functionalized graphene regions. High-energy states in F-SC have two special energy ranges including $-2.3<E/t<-1.9$ with holes localized only inside free graphene and $3<E/t<3.7$ with electrons localized only inside fluorinated graphene areas. Our analyses on real-space distributions of normalized probability density supply remarkable proof supporting the two characteristics. Calculated results of the fractal dimension of the conductance spectrum indicate that conductance fluctuations in H-SC and A-SC follow the Hausdorff fractal dimension behavior. Thus H-SC and F-SC indeed form the fractal dimension space. We also investigated the optical conductivity of H-SC and F-SC and find several conductivity peaks in high energy ranges as a result of adsorbed H or F atoms.

\begin{acknowledgements}
This work is supported by NSFC (Grant No. 11774269) and the Dutch Science Foundation NWO/FOM (Grant No. 16PR1024). X.Y., W.Z. and Q.Y. are supported by China Scholarship Council (CSC) under grant Nos. 202106270050, 202006270197 and 202006270212. Y.W. acknowledges the support from China Postdoctoral Science Foundation (No. 2019M660433), NSFC (No. 11832019) and NSAF (No. U1930402). Numerical calculations are operated in the Supercomputing Center of Wuhan University.
\end{acknowledgements}

\begin{appendix}

\section{Related parameters in TB model of F-SC} \label{sec:append_A}

%%%%%%%%%%%%%%%%%%%%%%%%%%%%%%%%% Table I %%%%%%%%%%%%%%%%%%%%%%%%
\begin{table}[H]
    \centering
    \caption {Slater-Koster matrix elements as a function of two-centre bond integrals.}
    \setlength{\tabcolsep}{12mm}{
    \begin{tabular}{c c}
    \hline
    $t_{ss}$&$V_{ss\sigma}$\\
    $t_{sp_{x}}$&$lV_{sp\sigma}$\\
    $t_{p_{x}p_{x}}$&$l^{2}V_{pp\sigma}+(1-l^{2})V_{pp\pi}$\\
    $t_{p_{x}p_{y}}$&$lm(V_{pp\sigma}-V_{pp\pi})$\\
    $t_{p_{x}p_{z}}$&$ln(V_{pp\sigma}-V_{pp\pi})$\\
    \hline
    \end{tabular}}
    \label{table1}
\end{table}
%%%%%%%%%%%%%%%%%%%%%%%%%%%%%%%%% Table I %%%%%%%%%%%%%%%%%%%%%%%%

%%%%%%%%%%%%%%%%%%%%%%%%%%%%%%%%% Table II %%%%%%%%%%%%%%%%%%%%%%%%
\begin{table}[H]
    \centering
    \caption{Direction cosines between diverse atom positions in graphene and fully fluorinated graphene (i.e., fluorographene). Here, $\delta_{F}$ is the vector of the adjacent upper F atom of the central C atom.}
    \setlength{\tabcolsep}{2.2mm}{
    \begin{tabular}{c c c c c c c}
    \hline\hline
    &  \multicolumn{3}{c}{graphene} & \multicolumn{3}{c}{fluorographene} \\
    \cline{2-4} \cline{5-7}
    $\delta_{i}$&$l_{i}$&$m_{i}$&$n_{i}$&$l_{i}$&$m_{i}$&$n_{i}$\\
    \hline
    $\delta_{1}$&$0$&$1$&$0$&$0$&$+0.964$&$-0.265$\\
    $\delta_{2}$&$-\sqrt{3}$/2&$-1/2$&$0$&$-0.835$&$-0.482$&$-0.265$\\
    $\delta_{3}$&$+\sqrt{3}/2$&$-1/2$&$0$&$+0.835$&$-0.482$&$-0.265$\\
    $\delta_{F}$&$0$&$0$&$0$&$0$&$0$&$1$\\
    \hline\hline
    \end{tabular}}
    \label{table2}
\end{table}

We extract the hopping and on-site potential parameters $t^{ij}_{\alpha\beta}$ and $\varepsilon^{i}_{\alpha}$ from $ab$ $initio$ calculation \cite{yuan2015electronic} by the Slater-Koster method \cite{slater1954simplified}. Table~\ref{table1} gives the matrix elements related to the two-center overlap integrals between diverse orbitals with involved direction cosines listed in Table~\ref{table2}. These hopping parameters for graphene and fluorographene are listed in Table~\ref{table3} \cite{yuan2015electronic}.

%%%%%%%%%%%%%%%%%%%%%%%%%%%%%%%%% Table III %%%%%%%%%%%%%%%%%%%%%%%%
\begin{table}[H]
    \centering
    \caption{Hopping and on-site potential parameters for graphene in the upper panel and for fluorographene in the lower panel. All these values are in unit of eV.}
    \setlength{\tabcolsep}{3mm}{
    \begin{tabular}{l c c}
    \hline\hline
    & graphene & fluorographene\\
    \hline
    $\varepsilon_{s}$&$-2.85$&$-5.54$\\
    $\varepsilon_{p_{xy}}$&$+3.20$&$+2.31$\\
    $\varepsilon_{p_{z}}$&$+0.00$&$+4.92$\\
    $V_{ss\sigma}$&$-5.34$&$-3.65$\\
    $V_{sp\sigma}$&$+6.40$&$+7.20$\\
    $V_{pp\sigma}$&$+7.65$&$+7.65$\\
    $V_{p_{xy}p_{z}\sigma}$&$+0.00$&$+2.20$\\
    $V_{p_{xy}p_{xy}\pi}$&$-2.80$&$-2.64$\\
    $V_{p_{xy}p_{z}\pi}$&$+0.00$&$-2.80$\\
    $V_{p_{z}p_{z}\pi}$&$-2.80$&$-1.87$\\
    \hline
    $\varepsilon^{F}_{p_{xy}}$&&$-4.94$\\
    $\varepsilon^{F}_{p_{z}}$&&$-1.69$\\
    $V^{C-F}_{sp\sigma}$&&$+1.06$\\
    $V^{C-F}_{pp\sigma}$&&$+9.85$\\
    $V^{C-F}_{pp\pi}$&&$-2.25$\\
    \hline\hline
    \end{tabular}}
    \label{table3}
\end{table}
%%%%%%%%%%%%%%%%%%%%%%%%%%%%%%%%% Table III %%%%%%%%%%%%%%%%%%%%%%%%

\section{Defect effects on the DOS of H-SC} \label{sec:append_B}

\begin{figure}[H]
\centering
\includegraphics[width=8.5cm]{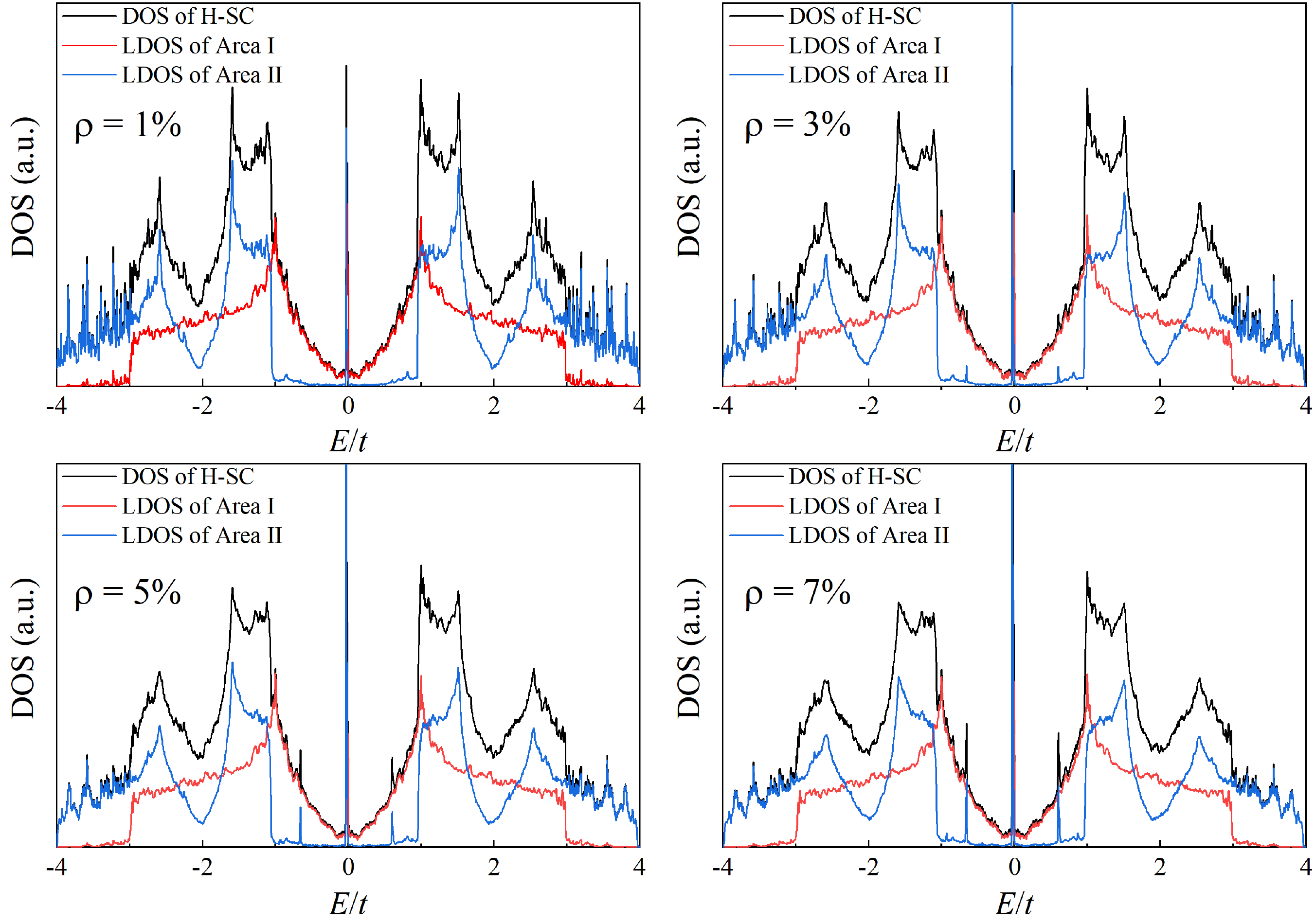}
    \caption{DOS of H-SC with different defect concentrations $\rho$. Here, $I=4$ and $W=297.5a$.} 
    \label{less_perfect_DOS}
\end{figure}

In realistic samples, less perfect adatom positions exist. We introduce the defect density $\rho$ measuring less perfect adatom positions, i.e., the percentage ratio between the number of random missed hydrogen atoms in functionalized Area II and the total number of hydrogen adatoms that ought to be in Area II. Figure~\ref{less_perfect_DOS} shows the LDOS and TDOS of H-SC with $I=4$ and $W=297.5a$ under different defect densities. For weak different defect densities with $\rho=1\%$ in Fig. \ref{less_perfect_DOS}(a) and $\rho=3\%$ in Fig. \ref{less_perfect_DOS}(b), the low-energy states inside $|E/t|\leq 1$ are still confined in Area I. However, under higher defect density in Figs. \ref{less_perfect_DOS}(c) and \ref{less_perfect_DOS}(d), the LDOS of Area II obviously increases at some energy ranges about $0.7\leq|E/t|\leq 1$. In this case, even for the low-energy states the H-SC with higher defect density can be not viewed an ideal SC fractal. 

%%%%%%%%%%%%%%%%%%%%%%%%%%%%%%%%% FIG 7 %%%%%%%%%%%%%%%%%%%%%%%%
%%%%%%%%%%%%%%%%%%%%%%%%%%%%%%%%% FIG 7 %%%%%%%%%%%%%%%%%%%%%%%%

\end{appendix}

\bibliographystyle{apsrev4-2}
\bibliography{ref}

\end{document}